\documentclass[iop,apj]{emulateapj} \usepackage{apjfonts} \newif\ifdraft \drafttrue \newif\ifpre \pretrue
\usepackage{graphicx}
\usepackage{xcolor}
\ifdraft
   \usepackage{hyperref}
   \hypersetup{
               colorlinks=true,
               linkcolor=black,
               filecolor=black,
               citecolor=black,
               urlcolor=blue
              }
     \newcommand{\web}[1]{\Blb{\url{#1}}}
\else
     \newcommand{\web}[1]{#1}
\fi

\bibpunct{(}{)}{;}{a}{}{,}
\newcommand{\PIMA}{$\cal P\hspace{-0.067em}I\hspace{-0.067em}M\hspace{-0.067em}A$ }
\newcommand{\Number}[1]{\ifnum#1<10\relax0\number#1\else\number#1\fi}
\newcommand{\isodate}{
\count151=\time
\divide\count151 by 60
\count151=\count151
\multiply\count151 by 60
\count152=\time
\advance\count152 by -\count151
\divide\count151 by 60
\count152=\count151
\multiply\count151 by 60
\count153=\time
\advance\count153 by -\count151
\Number{\year}.\Number{\month}.\Number{\day}--\Number{\count152}:\Number{\count153}
}
\renewcommand{\aa}{A\&A}
\newcommand{\ntab}[2]{ \multicolumn{1}{#1}{#2} }
\newcommand{\nntab}[2]{ \multicolumn{2}{#1}{#2} }
\newcommand{\nnntab}[2]{ \multicolumn{3}{#1}{#2} }

\definecolor{Dred}{rgb}{0.312,0.070,0.070}
\definecolor{Dblue}{rgb}{0.070,0.070,0.312}
\definecolor{Dgreen}{rgb}{0.070,0.312,0.070}
\definecolor{Db}{rgb}    {0.050,0.0,0.320}

\newcommand{\Blb}[1]{\textcolor{Dblue}{\bf #1}}

\newcounter{note}
\let\oldmarginpar\marginpar
\renewcommand\marginpar[1]{\-\oldmarginpar[\raggedleft\footnotesize #1]%
{\raggedright\footnotesize #1}}

\ifdraft
\fi
\begin{document}

\shorttitle{The position catalogue OBRS--1}
\title{The catalogue of positions of optically bright extragalactic
       radio sources OBRS--1}
\author{ L. Petrov}
\affil{ADNET Systems Inc./NASA GSFC, Greenbelt, MD 20771, USA}
\email{Leonid.Petrov@lpetrov.net}
\ifdraft
  \journalinfo{Astonomical Journal}
  \submitted{}
  \received{2011 June 14}
  \accepted{2011 July 21}
\fi

\begin{abstract}

It is expected that the European Space Agency mission Gaia will
make possible to determine coordinates in the optical domain of more
than 500\,000 quasars. In 2006, a radio astrometry project was 
launched with the overall goal to make comparison of coordinate systems 
derived from future space-born astrometry instruments with the coordinate 
system constructed from analysis of the global very long baseline 
interferometry (VLBI) more robust. Investigation of their 
rotation, zonal errors, and the non-alignment of the radio and optical 
positions caused by both radio and optical structures are important 
for validation of both techniques. In order to support these studies, 
the densification of the list of compact extragalactic objects that are 
bright in both radio and optical ranges is desirable. A set of 105 objects
from the list of 398 compact extragalactic radio sources with declination
$>-10\degr$ was observed with the VLBA+EVN with the primary goal of producing
their images with milliarcsecond resolution. These sources are brighter
than 18 magnitude at V band, and they were previously detected 
at the European VLBI network. In this paper coordinates of observed 
sources have been derived with milliarcsecond accuracies from analysis 
of these VLBI observations following the method of absolute astrometry. 
The catalogue of positions of 105 target sources is presented. The accuracies 
of sources coordinates are in the range of 0.3 to 7~mas, with the 
median 1.1~mas.

\end{abstract}

\keywords{astrometry --- catalogues --- surveys}

%

\maketitle

\section{Introduction}

   The method of Very Long Baseline Interferometry (VLBI) first proposed
by \citet{r:mat65} allows us to derive the position of sources with
nanoradian precision (1 nrad $\approx$ 0.2~mas). The first catalogue of
source coordinates determined with VLBI contained 35~objects
\citep{r:first-cat}. Since then, hundreds of sources have been observed under
geodesy and astrometry VLBI observing programs at 8.6 and 2.3~GHz
(X and S bands) using the Mark3 recording system at the International VLBI
Service for Geodesy and Astrometry (IVS) network. Analysis of these
observations resulted in the ICRF catalogue of 608~sources \citep{r:icrf98}.
Later, over 6000 sources were observed in the framework of the VLBA Calibrator
Survey (VCS) program \citep{r:vcs1,r:vcs2,r:vcs3,r:vcs4,r:vcs5,r:vcs6}, the VLBA
regular geodesy RDV program \citep{r:rdv}, the VLBA Imaging and 
Polarimetry Survey (VIPS) \citep{r:vips,r:astro_vips}, the VLBA Galactic plane 
Survey (VGaPS) \citep{r:vgaps}, the on-going Australian Long Baseline Array 
Calibrator Survey (LCS) \citep{r:lcs1}, and several other programs. 
The  number of extragalactic sources with positions determined from 
analysis of observations under absolute astrometry or geodesy programs 
reached 6455 by June 2011, and it continues to grow rapidly due 
to analysis of new observations and an on-going campaign of in depth 
re-analysis of old observations.

  The catalogue of positions of all these compact extragalactic radio sources
determined with VLBI\footnote{Available at \web{http://astrogeo.org/rfc}.}
with accuracies in a range of 0.05--30~mas forms a dense grid on the sky
that can be used for many applications, such as differential astrometry,
phase-referencing VLBI observations of weak objects, space navigation,
Earth orientation parameter determination, and space geodesy.
To date, this position catalogue is the most precise astrometric catalogue.
However, this high accuracy of positions of listed objects can be exploited
{\it directly} only by applications that utilize the VLBI technique.
Applications that use different observational techniques can benefit from
the high accuracy of VLBI positions only {\it indirectly} by observing common
objects from the VLBI catalogue with instruments at other wavelengths.
  For last three decades significant efforts were made for connecting
the VLBI position catalogue and existing optical catalogues made with
the use of ground instruments. An overview of the current status of
radio-optical connection and detailed analysis of the differences between
VLBI and optical source positions can be found in \citet{r:lfrq}.
According to them, the standard deviation of the differences between
the VLBI and optical catalogues is $\sim\!\!130$~mas. It was
shown by \citet{r:zah08} that when modern dedicated ground-based
observations are used, the differences are close to 30~mas.

  This level of agreement between VLBI and optical positions roughly
corresponds to the position accuracy of common objects from ground optical
catalogues, typically at a level of 100~mas. The European Space Agency
space-born astrometry mission Gaia, scheduled to be launched in 2013,
according to \citet{r:gaia} promises to reach sub-mas accuracies
of determining positions of quasars of 16--20 magnitude that will rival
accuracies of absolute astrometry VLBI. Since position catalogues produced
with Gaia and VLBI will be completely independent, their mutual rotations,
zonal differences and possibly other systematic effects can be interpreted
as errors of one of the techniques after resolving the differences
due to a misalignment of centers of optic and radio images of quasars
and a frequency-dependent core-shift \citep{r:kov08,r:por09,r:sokol11}.
Investigation of systematic differences will be very important for the
assessment of the overall quality of Gaia results and, possibly, the errors
in the VLBI position catalogue.

  This comparison will produce valuable results if 1)~it will be
limited to those common sources which VLBI positions are known with errors
smaller than several tenths of a milliarcsecond; 2)~the number of sources
will be large enough to derive meaningful statistics; and 3)~the sources
will be uniformly distributed over the sky. However, the number of quasars
that are a)~bright both in optical and radio wavelengths and therefore, can
be detected with both techniques (e.g. brighter than magnitude 18 as
suggested by \citet{r:mig03}) and b)~have a compact core, currently is
rather limited. Among 3946 radio sources with $\delta > -10\degr$
observed with the VLBA in the absolute  astronomy mode, 508 objects have
an association with a quasar or a BL~Lac object brighter than V~$18^m$
from the catalogue of \citet{r:vcv2010} within a $4''$ search radius.

   It was realized in mid 2000s that the densification of the list of such
objects is desirable. A specific program for identifying new VLBI
sources in the northern hemisphere, suitable for aligning the VLBI and Gaia
coordinate systems, was launched in 2006 \citep{r:bou08} with the eventual
goal of deriving highly accurate position of sufficiently radio-bright 
quasars from VLBI observations in the absolute astrometry mode. Since the 
current VLBI position catalogue is complete to the correlated flux density 
level of 200~mJy, the new candidate sources should necessarily be by a factor
of 2--4 weaker than that level. The original observing sample consisted
of 447 optically bright, relatively weak extragalactic radio sources with
declinations above $-10^{\circ}$. The detailed observing scheme of this
project is presented in \cite{r:bou08}. The first VLBI observations resulted
in the detection of 398 targets with the European VLBI Network (EVN)
\citep{r:bou10}, although no attempt to derive their positions of produce
images was made. VLBI observations of this sample in the absolute astrometry
mode promises to increase the number of optically bright radio sources with
precisely known positions by 80\%.

  As a next step of implementing this program, a subset of 105 detected
sources was observed with the global VLBI network that comprises the VLBA and
EVN observing stations with the goal of revealing their morphology on
milliarcsecond scales from VLBI images \citep{r:bou11} for consecutive
screening the objects with structure that potentially may cause non-negligible
systematic position errors. I present here results of astrometric analysis
from this VLBI experiment. Observations and their analysis are described in
sections \ref{s:obs} and \ref{s:anal}. The position catalogue is presented
in \ref{s:cat}. Concluding remarks are given in section \ref{s:summ}.

\section{Observations}
\label{s:obs}

  The observations used in this paper were carried out during a 48-hour
experiment GC030 on 7--9 March 2008 with a global VLBI array comprising
ten VLBA and 6 EVN stations ({\sc eflsberg}, {\sc hartrao},
{\sc medicina}, {\sc noto}, {\sc onsala60}, and  {\sc dss63} for part
of the time), simultaneously at S and X bands. The data were recorded
at 512~Mbps. The schedule was prepared by ensuring a minimum of three
5 minute long scans of each target source, while minimizing the slewing time
from source to source. In total, 115 objects, including 105 target sources
and 10 strong calibrators were observed during a 48-hour observing session.
Three target objects were observed in 2 scans, 20 target objects were
observed in 3 scans, 43 target objects were observed in 4 scans,
26 objects were observed in 5 scans, 10 objects were observed in 6
scans, 2 objects were observed in 7 scans, and 1 object was observed
in 8 scans. Antennas spent 78\% time recording signal from target sources.

  Although the overall goal of the observing program was absolute
astrometry, the design the GC030 experiment suffered several limitation and
was not favorable for determining sources coordinates with high accuracy.
First, the intermediate frequencies were selected to cover a continuous range
at both S and X-bands: 2.22699--2.29099~GHz and 8.37699--8.44099~GHz
respectively. There were two rationals behind selection that frequency setup
(P.~Charlot (2011), private communication). First, at the beginning of 2008,
the 512~Mbps mode was new. At that time, that setup was tested only for
a case of contiguously allocated intermediate frequencies (IFs). 
It was not clear whether every non-VLBA station will be able to support 
the wide-band mode. Since it happened in the past when a change in 
frequency setup ruined experiments,
it was decided to stay on the safe side and make the schedule using
contiguously spread IFs. Second, it was known (for example, D.~Gordon,
private communication, 2010) that AIPS implementation of fringe fitting,
task FRING, does not produce correct group delays when the IFs are spread
over the wide band. As a workaround, all absolute astrometry/geodesy
experiments prior 2010 were processed using a two-step approach: first the
fringe fit was made using data from each IF individually, and then
group delays over entire band were computed using fringe phases from each
individual IF derived in the previous step. The drawback of that approach
is that a source should be detected at each IF individually, which raises
the detection limit by $\sqrt{N}$, where $N$  is the number of IFs at
each band (4 in our case). Since the target sources were expected to be
weak, it was important to avoid a degradation of the detection limit by
a factor of 2. Work for developing an alternative fringe fitting
procedure \citep{r:vgaps}, free from this drawback was underway in 2008,
when the experiment was scheduled, but not finished at that time.
Unfortunately, group delays determined with the contiguously frequency
setup are {\it one order of magnitude} less precise with respect to the
frequency allocation traditionally used for absolute astrometry work
with the VLBA.

  The second limitation of the GC030 schedule for astrometry use
was a relatively rare observation of sources at low and high elevations for
better estimation of troposphere path delay in zenith direction. It was found
in the past that if to observe calibrator sources at low and high elevations
at each station every 1--2 hours, the reliability of estimates of the path
delay in the neutral atmosphere is significantly improved, and as a result,
systematic errors caused by mismodeling propagation effects are
reduced \citep{r:vcs3}.

  The third limitation of the GC030 schedule was a small number of sources
observed in prior astrometry/geodesy programs at dual S/X bands:
only 19 objects. Observations of a large number of sources, typically
30--60 objects in a 24 hour experiment, overlapping with previous
observations helps to establish firmly the orientation of the array
and to link positions of new sources with positions of other objects.

  Despite all these limitations, it was worth efforts to derive source
positions from such data since the a~priori positions of
these objects determined from Very Large Array (VLA)
observations \citep{r:first,r:nvss} were in the range of $0.03''$--$1''$.

\section{Data analysis}
\label{s:anal}

  The data were correlated at the Socorro hardware VLBA correlator.
The correlator computed the spectrum of cross correlation and
autocorrelation functions with frequency setup of 0.25~MHz at
accumulation intervals of 1.048576~s long.

  The procedure of further analysis is described in full details
in \citet{r:vgaps}. Here only a brief outline is given. At the first step,
the fringe amplitudes were corrected for the signal distortion in the sampler
and then calibrated according to measurements of system temperature and
elevation-dependent gain. Since the log files from VLBA sites for the second
half of the experiment were lost, no phase calibration was applied.
Then the group delay, phase delay rate, group delay rate, and fringe phase
were determined for all observations for each baseline at X and S bands
separately using the  wide-band fringe fitting procedure. These estimates
maximize the amplitude of the sum of the cross-correlation spectrum 
coherently averaged over
all accumulation periods of a scan and over all frequency channels in all IFs.
After the first run of fringe fitting, 12 observations at each baseline with
the strongest signal to noise ratios (SNR) were used to adjust the 
station-based complex bandpass corrections, and the procedure of computing 
group delays was repeated. This part of analysis is done with \PIMA\
software\footnote{Available at \web{http://astrogeo.org/pima}.}. Then the
results of fringe fitting were exported to the VTD/post-Solve VLBI analysis
software\footnote{Available at \web{http://astrogeo.org/vtd}.} for
interactive processing group delays with the SNR high enough to ensure
that the probability of false detection is less than 0.001. 
This SNR threshold is $5.8$ for the GC030 experiment. Detailed description of
the method for evaluation of the detection threshold can be found in
\citep{r:vgaps}. Then, theoretical path delays were computed
according to the state-of-the art parametric model as well as their partial
derivatives, and small differences between group delays and theoretical
path delay were used for estimation of corrections to a parametric model 
that describe the observations with least squares (LSQ). Coordinates
of target source, positions of all stations, except the reference one,
parameters of the spline that describes corrections to the a~priori path
delay in the neutral atmosphere in the zenith direction for all stations,
and parameters of another spline that describes the clock function
with the time span 1 hour were solved for in separate least square
solutions that used group delays at X and S bands individually.

  Observations that deviated by more than
$3.5\sigma$ in the preliminary solution were identified and temporarily
eliminated, and additive corrections to a~priori weights were determined.
The most common reason for an observation to be marked as an outlier
is a misidentification of the main maximum of the two-dimensional
Fourier-transform of the cross-spectrum. Then the fringe fitting procedure
was repeated for observations marked as outliers. But this time
the group delay and phase delay rate were evaluated for these observations
in a narrow window of 4~ns wide centered around the predicted value
of group delay computed using parameters of the VLBI model adjusted in
the preliminary LSQ solution. New estimates of group delays for points
with the probabilities of false detection less than 0.1, which corresponds
to the SNR $> 4.6$ for the narrow fringe search window, were used in the
next step of the interactive analysis procedure. The observations marked
as outliers in the preliminary solution and detected in the narrow window
at the second round of the fringe fitting were tried again.
If the new estimate of the residual was within 3.5 formal uncertainties,
the observation was restored and used in further analysis.
Parameter estimation, elimination of remaining outliers and adjustments
of additive weight corrections were then repeated. In total, 16629 matching
pairs of X and S band group delays out of 22750 scheduled were used in the
solution. Each source was detected at both bands and had the number
of dual-band pairs in the range of 19--321.

  The result of the interactive solution provided the clean dataset of
ionosphere-free linear combinations of X and S-band group delays with
updated weights. The dataset that was used for the final parameter estimation
utilized all dual-band S/X data acquired under absolute astrometry and space
geodesy programs from April 1980 through December 2010, including the data
from the GC030 experiment, in total 8 million observations. Thus,
the GC030 experiment was analyzed exactly the same way as over 5000 other
VLBI experiments, using the same analysis strategy that was used for
processing prior observations for ICRF, VCS, VGaPS, LCS, and K/Q survey
\citep{r:kq} catalogues. The estimated parameters are right ascensions and
declination of all sources, coordinates and velocities of all stations,
coefficients of B-spline expansion of non-linear motion for 17 stations,
coefficients of harmonic site position variations of 48 stations at 4 
frequencies: annual, semi-annual, diurnal, semi-diurnal, and axis offsets
for 67 stations. Estimated variables also included Earth orientation
parameters for each observing session, parameters of clock function and
residual atmosphere path delays in the zenith direction modeled with
the linear B-spline with interval 60 and 20 minutes respectively.
All parameters were adjusted in a single LSQ run.

  The system of LSQ equations has an incomplete rank and defines a family
of solutions. In order to pick a specific element from this family, I applied
the no-net rotation constraints on the positions of
212~sources marked as ``defining'' in the ICRF catalogue \citep{r:icrf98}
that required the positions of these sources in the new catalogue to have
no rotation with respect to their positions in the ICRF catalogue.
No-net rotation and no-net-translation constraints on site positions
and linear velocities were also applied. The specific choice of identifying
constraints was made to preserve the continuity of the new catalogue with
other VLBI solutions made during last 15 years.

  The global solution sets the orientation of the array with respect to
an ensemble of $\sim\!\!\!5000$ extragalactic remote radio sources.
The orientation is defined by the continuous series of Earth orientation
parameters and parameters of the empirical model of site position
variations over 30 years evaluated together with source coordinates.
Common sources observed in the GC030 experiment as amplitude calibrators
provided a connection between the new catalogue and the old catalogue of
compact sources.

  As a valuable by-product of GC030 observations, positions of {\sc dss63}
station were determined (see Table~\ref{t:dss63}). To my knowledge, 
this is the only S/X experiment with participation of this station that 
can be found in publicly accessible databases. Velocity of {\sc dss63} 
was constrained to be the same as velocity of {\sc dss65} station that 
is located in 1440 meters from {\sc dss63}.

  Radio images of observed sources in both S and X bands were presented 
in a graphical form in \citet{r:bou11}. In order to provide 
a measure of source strengths at long and short baselines for predicting 
the SNR in future observations, I made my own simplified amplitude analysis 
and derived the median correlated flux densities at baseline projection 
lengths shorter than 900~km and longer than 5000 km. This procedure is 
described in details in \citet{r:lcs1}. It is outlined here briefly. First, 
I computed the a~priori system equivalent flux density (SEFD) using system 
temperatures and gain curves for each antenna. Fringe amplitudes for every 
observation used in astrometric analysis, except those marked as outliers,
were converted to flux densities by multiplying them by the square root
of the product of the a~priori SEFDs of both stations of a baseline.
Then I adjusted multiplicative gain corrections from logarithms of ratios
of observed correlated flux densities of 10 amplitude calibrators to
their values predicted on the basis of publicly available brightness
distributions\footnote{Available at \web{http://astrogeo.org/vlbi\_images}}
using least squares. I applied these corrections to estimates of correlated
flux densities of observed sources and computed the median value at two
ranges of baseline projections. Comparison of estimates of median
correlated flux densities derived by this method with estimates of
correlated flux densities generated using images produced by a rigorous
self-calibration procedure for three 24~hour survey experiments VCS5
\citep{r:vcs5} showed that the accuracy of median correlated flux densities
estimated using the simplified method is at a level of 15\%.
These estimates of correlated flux densities are complementary to
source image statistics shown in \citet{r:bou11}, for instance, to their
total flux densities integrated from X- and S-band images.

\begin{table}[hb]
   \caption{Position of {\sc dss63} station at the 2000.0 epoch determined from
            the GC030 experiment. Its velocities listed in the right column
            were constrained to be the same as velocities of station
            {\sc dss65}.}
   \label{t:dss63}
   \renewcommand{\arraystretch}{1.2}
   \begin{tabular}{l @{\quad} r @{\quad} r}
             & Position (m) & Velocity (mm/yr) \\
          \hline
          X & $  4849092.429 \pm 0.025 $ & $ -1.63 \pm 0.23 $ \\
          Y & $ -3601804.438 \pm 0.013 $ & $ 18.49 \pm 0.09 $ \\
          Z & $  4115109.146 \pm 0.024 $ & $ 10.62 \pm 0.24 $ \\
   \end{tabular}
\end{table}

\section{The catalogue}
\label{s:cat}

\begin{table*}[t]
   \caption{First 12 rows of the OBRS--1 source position catalogue.}
   \label{t:cat}
   \begin{tabular}{ l l r r r r r r r r r r @{\enskip} r}
      \hline
      \nntab{c}{IAU name} &
      \nntab{c}{Source coordinates} &
      \nnntab{c}{Position errors} &
       &
      \nntab{c}{$F_{corr}$ S-band} &
      \nntab{c}{$F_{corr}$ X-band} &
      Flag \\
      B1950  &
      J2000  &
      \ntab{c}{$ \alpha $ } &
      \ntab{c}{$ \delta $ } &
      $ \sigma_\alpha $ &
      $ \sigma_ \delta $ &
      Corr &
      \#pnt &
      short &
      unres &
      short &
      unres &
      \\
        & &
        \ntab{l}{~hr~mn~sec} &
        \ntab{l}{~~~~$^{\circ}$~~~~$^\prime$~~~~~$^{\prime\prime}$}  &
        \ntab{c}{mas} & \ntab{c}{mas} & & &
        \ntab{c}{Jy}  & \ntab{c}{Jy}  &
        \ntab{c}{Jy}  & \ntab{c}{Jy}  & \\
      \hline
      0003$+$123 & J0006$+$1235 & 00 06 23.056086 & $+$12 35 53.09833 &  0.26 &  0.42 & $ 0.107$ &  198 &  0.137 & 0.071 & 0.138 & 0.083 & X \\
      0049$+$003 & J0052$+$0035 & 00 52 05.568998 & $+$00 35 38.14614 &  0.89 &  3.16 & $-0.631$ &   32 &  0.031 & 0.024 & 0.062 & 0.061 &   \\
      0107$-$025 & J0110$-$0219 & 01 10 13.160493 & $-$02 19 52.84055 &  0.65 &  2.26 & $-0.476$ &   98 &  0.074 & 0.069 & 0.061 & 0.058 &   \\
      0109$+$200 & J0112$+$2020 & 01 12 10.190819 & $+$20 20 21.76438 &  0.31 &  0.67 & $-0.305$ &  170 &  0.095 & 0.070 & 0.125 & 0.087 &   \\
      0130$-$083 & J0132$-$0804 & 01 32 41.126050 & $-$08 04 04.83517 &  1.38 &  1.57 & $-0.029$ &   64 &  0.104 & 0.083 & 0.065 & 0.058 &   \\
      0145$+$210 & J0147$+$2115 & 01 47 53.822855 & $+$21 15 39.72637 &  0.40 &  1.15 & $-0.413$ &  127 &  0.295 & 0.210 & 0.107 & 0.053 &   \\
      0150$+$015 & J0152$+$0147 & 01 52 39.610907 & $+$01 47 17.38264 &  0.86 &  2.75 & $-0.576$ &   70 &  0.046 & 0.047 & 0.048 & 0.042 &   \\
      0210$+$515 & J0214$+$5144 & 02 14 17.934429 & $+$51 44 51.94772 &  0.39 &  0.36 & $-0.064$ &  401 &  0.092 & 0.076 & 0.059 & 0.046 & X \\
      0446$+$074 & J0449$+$0729 & 04 49 21.170617 & $+$07 29 10.69568 &  0.63 &  1.31 & $-0.147$ &   80 &  0.043 & 0.036 & 0.092 & 0.074 &   \\
      0502$+$041 & J0505$+$0415 & 05 05 34.769151 & $+$04 15 54.57316 &  2.02 &  5.40 & $-0.381$ &   41 &  0.036 & 0.051 & 0.032 & 0.036 &   \\
      0519$-$074 & J0522$-$0725 & 05 22 23.196279 & $-$07 25 13.47580 &  4.96 &  7.17 & $ 0.088$ &   19 &  0.043 & 0.043 & 0.053 & 0.038 &   \\
      0651$+$428 & J0654$+$4247 & 06 54 43.525947 & $+$42 47 58.73588 &  0.48 &  0.64 & $ 0.494$ &  168 &  0.098 & 0.101 & 0.075 & 0.077 &   \\
      \hline
   \end{tabular}
   \tablecomments{Table~\ref{t:cat} is presented in its entirety 
                  in the electronic edition of the Astronomical Journal. 
                  A portion is shown here for guidance regarding 
                  its form and contents.
                 }
\end{table*}

I have determined positions of 105 sources observed in GC030 experiment.
They are listed in Table~\ref{t:cat}. Although positions of all 5336 
astrometric sources were adjusted in the LSQ solution that included the 
OBRS--1 sources, only coordinates of the 105 target sources observed during 
GC030 experiment are presented in the table. The 1st and 2nd columns give 
the IVS source name (B1950 notation) and IAU name (J2000 notation). 
The 3rd and 4th columns give source coordinates at the equinox on 
the J2000.0 epoch. Columns 5 and 6 give formal source position uncertainties 
in right ascension and declination in mas (without $\cos\delta$ factor), 
and column 7 gives the correlation coefficient between the errors in right 
ascension and declination. The number of group delays used for position 
determination is listed in column 8. Columns 9 and 10 provide the median 
value of the correlated flux density in Jansky at S-band at baseline 
projection lengths shorter than 900~km and at baseline projection lengths 
longer than 5000~km. The latter estimate serves as a measure of the 
correlated flux density of the unresolved component of a source. 
Columns 11 and 12 provide the median of the correlated flux density at 
X-band at baselines shorter than 900~km and longer than 5000~km. The last 
column contains a cross-reference flag: V if a sources was observed in 
VIPS campaign, X if it was observed at X/S bands in other absolute 
astrometry campaign, and VX if it was observed in both.

\begin{figure}[tbh]
   \includegraphics[width=0.48\textwidth,clip]{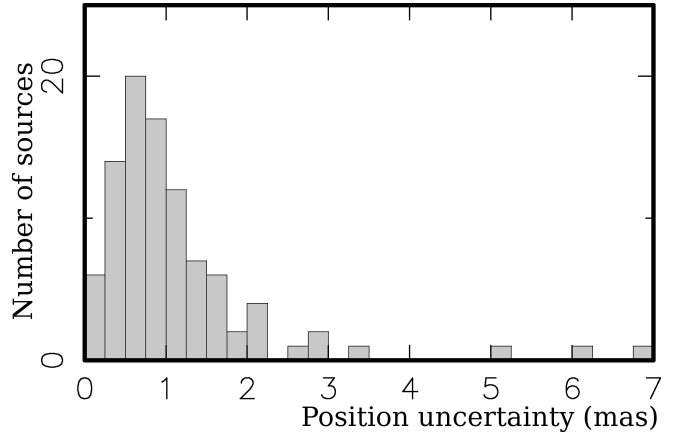}
   \caption{The histogram of the semi-major axes of position
            error ellipses among 105 target sources in the OBRS--1
            catalogue.
   \label{f:obrs1_hist}
   }
\end{figure}

  Uncertainties in sources position that were observed only in the GC030
experiment range from 0.3~mas (\object{1345+735})
to 7.2~mas (\object{0519-074}) with the median 1.1~mas. The distribution
of the semi-major axes of position error ellipses is presented in
Figure~\ref{f:obrs1_hist}.

\section{Error analysis}

  Among 105 target sources, 26 objects were observed in VIPS C-band
(5~GHz) program and 9 objects were observed in dual-frequency S/X VLBA
experiments under absolute astrometry programs. For comparison
purposes, I made a trial solution that used exactly the same
setup as the main solution, but excluded 9 common objects from the
GC030 experiment. The results of the comparison presented
in Table~\ref{t:diff} shows that, except for
declination of \object{2043+749}, the differences are within formal
uncertainties of the OBRS--1 catalogue. The formal uncertainties of
source positions are computed from standard deviations of group delay
estimates using the law of error propagation. Since the selection of
intermediate frequencies was unfavorable for a precise determination of
group delays and the sources were relatively weak, the thermal noise
dominates the error budget.

  Realistic uncertainties of parameter adjustments can be evaluated
only by exploiting some redundancy in the data or by using additional
information. We do not have enough redundancy to evaluate rigorously the
level of systematic errors in the GC030 campaign. Comparison of positions
of 9~common sources indicates that systematic errors, if exist, do not
exceed 1~mas. Although only 10 atmosphere calibrators were included in
the schedule, 3--5 times less than in dedicated absolute astrometry observing
experiments, they were observed rather intensively. We can indirectly
estimate the level of systematic errors caused by the sparseness of the
distribution of calibrator sources by comparing the source distribution
in experiment GC030 with that in the prior VLBA Calibrator Survey program
VCS1 \citep{r:vcs1}. The azimuthal-elevation distribution of sources observed
in that campaign for a central VLBA antenna (Figure~3b in \citet{r:vcs1}) was
concentrated in a narrow band at the sky for 95\% of the sources, and
very few atmospheric calibrators outside that band were used. The reliability
of estimation of atmosphere path delays in zenith direction was significantly
compromised, and as a result, the formal uncertainties from the LSQ solution
had to be inflated by adding in quadrature the error floor
of 0.4~mas.

  The VCS1 campaign can be considered as an extreme case of the
effect of the non-uniform distribution of observed sources. Both calibrator
sources and targets in GC030 were distributed more uniformly than in the VCS1
campaign. Analysis of estimates of residual atmosphere path delays does
not show abnormalities. I surmise tentatively that systematic
errors of the OBRS--1 catalogue are probably do not exceed 0.4~mas, which
is insignificant with respect to its random errors. I presented formal
uncertainties from the LSQ solution ``as is'', leaving investigation of
systematic errors in depth in the future when more observations in this
mode will be collected.


\begin{table*}[ht]
   \caption{Differences between estimates of coordinates of 9 common target
            sources determined using only GC030 observations and estimates from
            other X/S VLBA absolute astrometry experiments.}
   \label{t:diff}
   \begin{tabular}{ l l r r r r r r r r}
       \hline
       \nntab{c}{Source name} & \nntab{c}{Position difference} &
       \nntab{c}{XS source position} & \nnntab{c}{X/S position uncertainty} &
       \phantom{$\bigl(\bigr)$} \\
      B1950-name & J2000-name   & $ \Delta \alpha \cos \delta $ &
       \ntab{c}{$ \Delta\delta $} & Right ascension & Declination &
        $ \sigma(\alpha)$ & $ \sigma(\delta)$ & Corr & \# Obs \\
        & & \ntab{c}{mas} & \ntab{c}{mas} & \ntab{l}{~hr~mn~sec}  &
          \ntab{l}{~~~~$^{\circ}$~~~~$^\prime$~~~~~$^{\prime\prime}$}  &
          \ntab{c}{mas} & \ntab{c}{mas} & & \\
      \hline
      0003$+$123 & J0006$+$1235 & $ -0.4 \pm 0.6 $ & $  0.0 \pm 0.8 $ & 00 06 23.05607 & +12 35 53.0983 & 0.3 & 0.5 &  0.124 &   62 \\
      0210$+$515 & J0214$+$5144 & $ -1.3 \pm 0.6 $ & $ -0.1 \pm 0.8 $ & 02 14 17.93433 & +51 44 51.9475 & 0.8 & 0.5 &  0.296 &   99 \\
      0708$+$742 & J0714$+$7408 & $ -0.3 \pm 0.4 $ & $ -0.2 \pm 0.4 $ & 07 14 36.12502 & +74 08 10.1440 & 0.6 & 0.2 &  0.085 &   88 \\
      1721$+$343 & J1723$+$3417 & $  0.3 \pm 0.6 $ & $  0.3 \pm 1.1 $ & 17 23 20.79594 & +34 17 57.9652 & 0.2 & 0.4 & -0.432 &   76 \\
      1759$+$756 & J1757$+$7539 & $  0.8 \pm 0.8 $ & $  2.8 \pm 1.6 $ & 17 57 46.35883 & +75 39 16.1800 & 1.3 & 0.4 &  0.375 &  276 \\
      2043$+$749 & J2042$+$7508 & $ -0.6 \pm 0.6 $ & $ -3.7 \pm 1.2 $ & 20 42 37.30776 & +75 08 02.4415 & 1.4 & 1.1 &  0.211 &   45 \\
      2111$+$801 & J2109$+$8021 & $  0.6 \pm 1.7 $ & $ -0.6 \pm 2.2 $ & 21 09 19.16511 & +80 21 11.2264 & 9.8 & 2.2 & -0.032 &   13 \\
      2316$+$238 & J2318$+$2404 & $  0.3 \pm 0.3 $ & $  0.1 \pm 0.7 $ & 23 18 33.96785 & +24 04 39.7496 & 0.3 & 0.4 &  0.050 &   72 \\
      2322$+$396 & J2325$+$3957 & $  0.3 \pm 0.4 $ & $ -0.5 \pm 0.8 $ & 23 25 17.86983 & +39 57 36.5084 & 0.2 & 0.2 & -0.274 &  118 \\
      \hline
   \end{tabular}
\end{table*}

\section{Discussion}

  In the course of development of radio astrometry for last 40 years,
we learned that in order to derive precise source positions using the method
of absolute astrometry, a VLBI experiment should 1)~have intermediate 
frequencies spread as wide as possible over the band(s); 2)~observe every
1--2~hours blocks of 3--5 sources with at least one source at
elevations $20\degr$ above the horizon and one source at elevations $55\degr$
above the horizon; 3)~collect enough bits for detection target sources
at long baselines.

  Unfortunately, the selection of intermediate frequencies in the GC030
experiment did not satisfy the first condition. The choice of intermediary
frequencies is not very important for producing source images and observers
often record a continuous bandwidth. But this choice is critical for absolute
astrometry applications, since precision of group delay is reciprocal to the
variance of the frequencies in the band. The choice is especially important
for astrometry of weak sources, since unlike to observations of bright
sources when systematic errors dominate the error budget, the position
accuracy of weak sources is determined by the uncertainties of group delays
caused by the thermal noise. The frequency setup spread over 494~MHz used
in VLBA geodesy/astrometry RDV program \citep{r:rdv} had the uncertainties 
of group delay by a factor of 11.1 smaller than in the GC030 experiment at 
a given signal to noise ratio (SNR). The VLBA hardware allows to spread the 
IFs over 1000~MHz that brings uncertainties of group delay down even further
by a factor of~2 \citep{r:wide-memo11}.

  It should be stressed that there is no necessity to limit the spread of
intermediate frequencies for image experiments. One of the most extensive
dedicated imaging program, the VLBI Image and Polarization Survey (VIPS)
\citep{r:vips} used 4 IFs spread over 494~MHz in order to improve the
$uv$ coverage and to allow for rotation measure determinations \citep{r:tay05}.
Analysis of both VIPS and RDV observations provided excellent source
maps \citep{r:vips,r:rdv_astro,r:pus08}. Maps from absolute astrometry
observations typically have dynamic range 1:100--1:1000 (see \citet{r:vcs6}
and references therein). These maps allowed \citet{r:cha07} to determine
source structure indexes and make conclusions about suitability of sources
for precise astrometry.

  The approach proposed by \citet{r:bou08} to run 3 observing campaigns
for an absolute astrometry program, first for detection, second for producing
source maps, third for deriving source positions deviates sharply from
the strategy used for last 40 years for determining positions of 6000 sources,
which used one campaign per program.

  Our analysis of GC030 experiment shows that running 2 separate observing
campaigns for imaging and astrometry, which doubles requested observing time,
is not the best choice. Spreading the intermediate frequencies over 500~MHz
would reduce random errors of position estimates by a factor of 11, i.e. the
median position error would be 0.1~mas, without compromising imaging results.
With such precise group delays, position accuracy would be limited by
systematic errors. More intensive observations of troposphere calibrators
for mitigation systematic errors would require approximately 5--8\% additional
observing time according to \citet{r:vgaps}. That means that the goal of
the project could be reached by using 2 runs instead of 3, which
requires one half of requested resources.

  Including {\sc eflsberg} in the array is beneficial, because this
station improves the baseline sensitivity at X-band by a factor of 4,
which is important for detecting weak sources. The benefit of using other
European stations and especially a station in South Africa which has almost
no mutual visibility with both American and European stations is less obvious.
In order to access the impact of other stations on the source position
estimates, I made a trial solution that excluded {\sc medicina},
{\sc hartrao}, {\sc noto}, {\sc onsala60}, and  {\sc dss63} from GC030.
Comparison of position differences showed that they are within formal
uncertainties. An average increase of uncertainties of the trial solution
using the data from the restricted array was 20\% for right ascensions and
30\% for declinations. The median increase was 28\% and 42\% respectively.
Removing EVN stations from the array would, of course, degrade the quality
of images, but as analysis of other VLBA experiment showed, for instance
K/Q survey \citep{r:kq}, not to the level that would undermine their
usability for the goals of this specific project.

  These are important lessons that we learned from analysis of
these observations.

\section{Summary}
\label{s:summ}

   Analysis of the first dual-band S/X VLBA experiment of the campaign for
observing optically bright extragalactic radio sources allowed us to
determine positions of 105 target sources. Despite using the frequency setup
unfavorable for absolute astrometry, the position uncertainties ranged
from 0.3 to 7~mas with the median value of 1.1~mas. The sources were
relatively weak: the median correlated flux density at baselines longer
than 5000~km ranged from 25 to 190~mJy with the median value around 60~mJy
at both bands, which is a factor of 2 weaker than in the VLBA Calibrator
surveys. However, recording at 512~Mbps with integration length of 300~s
was sufficient to detect 73\% of the observations, including those 
at long baselines.

  A position accuracy of 1~mas is sufficient for using these 
sources as phase calibrators, but not sufficient for drawing meaningful 
conclusions from comparison of Gaia and VLBI positions. All the sources 
will have to be re-observed with the wide-band frequency setup in order 
to reach 0.1~mas level of accuracy.

  In 2010--2011, the remaining 293 sources were observed 
at the VLBA + EVN. These observations will help us to further extend 
the position catalogue of optically bright radio sources.

\acknowledgements

It is my pleasure to thank G\`{e}raldine Bourda and Patric Charlot
for fruitful discussions. The National Radio Astronomy Observatory is
a facility of the National Science Foundation operated under cooperative
agreement by Associated Universities, Inc.

{\it Facilities:} \facility{VLBA (project code GC030)}.

\end{document}